\documentstyle[aps,epsfig]{revtex}
 \def\ket{\rangle}
\begin{document}
 \title{Theoretical efficient high capacity Quantum Key Distribution Scheme}
 \author{ G. L. Long$^{1,2,3,4}$ and X. S. Liu$^{1, 5}$}
 \address{$^{1}$Department of Physics, Tsinghua University, Beijing 100084, P.R.
China\\
$^2$ Key Laboratory for Quantum Information and Measurements, Ministry of Education, %%@
Beijing 100084,  P.R. China \\
$^3$ Institute of Theoretical Physics, Chinese Academy of Sciences, Beijing 100080, P. R. %%@
China\\
$^4$ Center for Atomic, Molecular and NanoSciences, Tsinghua University, Beijing 100084, P %%@
R China\\
$^5$ Department of Physics, Shandong Normal University, Jinan 250015, P. R. China}
\maketitle
\date{\today}
\begin{abstract}
A theoretical quantum key distribution scheme using EPR pairs is presented.  This scheme %%@
is efficient in that it uses all EPR pairs in distributing the key except those chosen for %%@
checking eavesdroppers. The high capacity is achieved because each EPR pair carries 2 bits %%@
of key code. 
\end{abstract}

Since languages become the tool for communication, the desire and need to transmit secret %%@
messages from one person to another begin.  Then human have the cryptography -- an art to %%@
transmit information so that it is unintelligible and therefore useless to those who are %%@
not meant to have access to it. The most important classic cryptographic scheme is %%@
public-key crypto-system\cite{r1}, its safety relies on the high complexity of the %%@
underlying mathematical problems, for instance the factorization of large numbers. But %%@
with the development of the quantum computation(QC), especially the Shor's algorithm for  %%@
factoring  big numbers, the systems once seemingly unbroken in practice will be agressed %%@
easily. Now in the information  community, the safety of transmission of secret %%@
information is becoming more and more concerned. One essential theme of secure %%@
communication is to distribute secret keys  between senders and  receivers. Quantum %%@
mechanics, one of the greatest discovery of the 20th century, has now entered the field of %%@
cryptography: if the key distribution makes use of quantum states, an eavesdropper can not %%@
measure them without disturbing them. The principle of the quantum mechanics can help to %%@
make the key distribution secure. Upto now, there have already been several quantum key %%@
distribution(QKD) schemes: BB84 protocol\cite{r3},  the EPR %%@
scheme\cite{r5,r5a},B92\cite{r4}, the 4+2 protocol\cite{r5p}, the six-state %%@
protocol\cite{r5p2}, the Goldenberg/Vaidman scheme\cite{r5p3}, Koashi/Imoto %%@
scheme\cite{r5p4}, and the recent  Cabello protocol\cite{rcabello} and so on. 

Experimental research on QKD is progressing very fast, for instance the optical-fiber %%@
experiment of BB84 and B92 protocols have been realized upto 48 km\cite{r5p5} and %%@
experiment in free space of  B92 scheme has been achieved over 1 km distance\cite{r5p6}, %%@
and very recently  upto 1.6 km during daylight\cite{r5p7} .
 
Before presenting our scheme, we first introduce the notations. An EPR pair is one of the %%@
4 Bell states
\begin{eqnarray}
              \mid\psi_1\rangle={1\over \sqrt{2}}(\mid00\rangle+\mid11\rangle)\nonumber\\
              \mid\psi_2\rangle={1\over \sqrt{2}}(\mid00\rangle-\mid11\rangle)\nonumber\\
              \mid\psi_3\rangle={1\over \sqrt{2}}(\mid10\rangle+\mid01\rangle)\nonumber\\
              \mid\psi_4\rangle={1\over \sqrt{2}}(\mid10\rangle-\mid01)\rangle.
              \label{e1}
\end{eqnarray}
Alice and Bob agree beforehand that $\mid\psi_1\rangle$, $\mid\psi_2\rangle$, %%@
$\mid\psi_3\rangle$, $\mid\psi_4\rangle$ are encoded as  00,01,10,11 respectively. This %%@
coding increases the capacity of our scheme. An ordered  $N$  EPR particle pair sequence %%@
is denoted by  $[(P_1(1), P_1(2)), (P_2(1), P_2(2))$,  $...$,  $ (P_i(1), P_i(2)),..., %%@
(P_N(1), P_N(2))]$.  We denote $P_i(1)$ for one particle in the i-th EPR pair, and %%@
$P_i(2)$ for the other, and $i=1, 2, \ldots, N$. We say that $P_i(1)$ is the EPR partner %%@
particle of  $P_i(2)$ and vice versa. The order of these $N$ EPR pairs is maintained %%@
through out the QKD process.  We can also take one EPR partner particle $P_i(1)$ from each %%@
EPR pair $(P_i(1), P_i(2))$ to form an EPR partner particle sequence $[P_1(1), P_2(1),$ $ %%@
..., P_i(1), ..., P_N(1)]$. A Bell-basis measurement  is joint measurement of two %%@
particles onto the 4 Bell basis states.  

Our protocol is as follows: 

 (1) Alice produces an ordered $N$ EPR pair sequence, $[(P_1(1), $ $P_1(2)),$ $ (P_2(1), %%@
P_2(2)), ...,$ $ (P_i(1), P_i(2)),$ $..., $ $(P_N(1), P_N(2))]$. 

 (2) Then Alice takes one particle from each EPR pair to form an ordered EPR partner %%@
particle sequence: $[P_1(2)$, $P_2(2)$, $P_3(2)$, ...,$P_N(2)]$. The rest of the EPR %%@
partner particles form another ordered EPR partner particle sequence: $[P_1(1)$, $P_2(1)$, %%@
$P_3(1)$, ...,$P_N(1)]$. Alice sends to Bob one ordered EPR partner particle sequence: %%@
$[P_1(2)$, $P_2(2)$, $P_3(2)$, ...,$P_N(2)]$.  

(3) After Bob receives the ordered EPR partner particle sequence, he chooses randomly %%@
among the EPR partner set a sufficiently large subset and performs measurement on the %%@
particles in the subset. The result of this measurement will be either 0 or 1. Bob stores %%@
the rest of the particles of his EPR particle sequence. 

(4) Then Bob tells Alice through a classical channel (such as a telephone line) that he %%@
has received the particle sequence, and the particles that he has chosen to measure in a %%@
some direction. After hearing from Bob,  Alice then performs measurement on the partner %%@
subset of those particles whose partner has been measured by Bob. Alice and Bob then %%@
publicly compare the results of these measurement to check evesdropping. We refer this %%@
procedure as the first evesdropping check. 

(5) If they are certain   that there is no evesdropping, then Alice sends Bob the %%@
remaining EPR particle sequence: $[P_1(1)$, $P_2(1)$, $P_3(1)$, ...,$P_N(1)]$. Of course, %%@
the particles that have been measured are dropped from this particle sequence. 

(6) After Bob receives these $N$ particles, he takes one particle from each particle %%@
sequence in order  and performs Bell-basis measurement on them. He records the results of %%@
the  measurements. 

(7) Alice and Bob choose a sufficiently large subset of these Bell-basis measurement %%@
results to check if the QKD is successful. If the error rate in this check is below  %%@
certain threshod, then the Bell-basis measurement  results are taken as  raw keys. We call %%@
this procedure as the second evesdropping check. 

 The procedure is shown in Fig. 1.  During the transmission of the second particle %%@
sequence, Eve can not get access to the EPR pairs, hence can not steal the key. Her action %%@
just causes disturbance to the key, which in fact is a kind of destruction. The second %%@
evesdropping check is designed for detecting this. In practice, this procedure may well be %%@
combined with  the privacy amplification procedure in the post-processing stage of QKD. %%@
Next we discuss the security of the protocol.

First, the scheme is secure against direct measurement by Eve. In this attack, Eve %%@
intercepts the first particle sequence and makes measurements on them, then she resends %%@
these measured particle sequence to Bob. Because of Eve's measurement, all the EPR pairs, %%@
with one particle sequence at Alice's hand and the other particle set at Bob's hand,  are %%@
destroyed. During the first evesdropping checking procedure, Eve's destruction is not %%@
detectable because Bob's measurement will yields exactly the same results as Eve which is %%@
consistent with the results of Alice's measurement. However during the second evesdropping %%@
check, Eve's action is easily detected.   Because the EPR pairs have collapsed, Bob will %%@
have only 50\% probability of obtaining the right result when Bob uses Bell-basis %%@
measurement to ``read''  his ``EPR" particle pairs. For instance, suppose %%@
$|00\ket+|11\ket$ is collapsed into $|00\ket$ by Eve's interception. Since
\begin{displaymath}
|00\ket={1\over \sqrt{2}}|\psi_1\ket+{1\over \sqrt{2}}|\psi_2\ket,
\end{displaymath}
Bob has only 50\% to obtain $|\psi_1\ket$ when he makes a Bell-basis measurement. In order %%@
words, the error rate will be as high as 50\%, and this can be easily detected. Eve can %%@
not obtain any useful information from this destructive attack.

Secondly, the scheme is secure against the intercept-resend attack.  Suppose  Eve %%@
intercepts  the particle sequence $\{P_i(2), i=1,...,N\}$ and  keeps them. 
However she can not make Bell-basis measurement because she does not possess the other %%@
particle sequence. In order to obtain the other particle sequence, she must send a fake %%@
particle sequence to Bob so that Bob can notify Alice. The particle sequence sent by Eve %%@
to Bob may well be a particle sequence from  an EPR pair sequence $[(P_1^*(1), P_1^*(2)), %%@
(P_2^*(1), P_2^*(2)), ...,$ $ (P_i^*(1), P_i^*(2)),..., (P_N^*(1), P_N^*(2))]$. However %%@
this can be detected easily during the first evesdropping check. Bob chooses randomly a %%@
subset of particles and measures them. After Bob tells Alice what particles he has %%@
measured, Alice measures the corresponding particles at her hands. Then Alice and Bob %%@
publicly compare their results.  If Bob's particle sequence is the fake particle sequence %%@
sent by Eve, half of his results during the first evesdropping check will be inconsistent %%@
with that of Alice. This will easily detect Eve. We can see this more clearly by studying %%@
the mutual information defined as\cite{rcabello} 
\begin{displaymath}
I(X:Y)=H(X)-H(X|Y),
\end{displaymath}
where $H(X)=-\sum_i p(x_i)\log_2 p(x_i)$  is the Shannon entropy, which is a function of %%@
the probabilities $p(x_i)$ of all possible values of $X$, and  the sum is over those $i$ %%@
with $p(x_i)>0$. $H(X|Y)$ is the expected entropy of $X$ once one knows the value of $Y$, %%@
and is given by
\begin{displaymath}
H(X|Y)=\sum_j p(y_j)\left[-\sum_i p(x_i|y_j)\log_2 p(x_i|y_j)\right].
\end{displaymath}
If there is no evesdroping, the mutual information between Alice and Bob is $I_{AB}=2$, %%@
and the mutual information between Alice and Eve is zero. When there is evedropping, the %%@
mutual information between Alice and Eve is $I_{AE}=2$, and the mutual information between %%@
Alice and Bob is $I_{AB}=0$. Evedropping in this scheme is easily detected. This is %%@
compared with the BB84 protocol in which Eve evesdrops with the same method as Bob,  %%@
$I_{AB}=5/8 \log_2 5+ 3/8 \log_2 3-2\doteq 0.046$, and $I_{AE}=I_{EB}=3/4 \log_2 3-1\doteq %%@
0.189$. When there is no evedropping, the mutual information between Alice and Bob in the %%@
BB84 scheme is $I_{AB}=3/4 \log_2 3-1\doteq 0.189$.

Thirdly, the scheme is safe against the opaque attack strategy. In this strategy Eve %%@
intercepts every signal and measures them. If she gets a result, she just let the signal %%@
go. Otherwise she destroys the signal completely. In our scheme Eve can not use this %%@
strategy, because Alice sends to Bob  only one particle sequence at a time. This particle %%@
sequence  is useless without the other particle sequence. If Eve tries to hide something, %%@
the QKD process simply stopped.

Like other QKD protocols using orthogonal states, one distinct feature of our scheme is %%@
its high efficiency in terms of number of keys sent to the number of EPR pairs(particles) %%@
used\cite{rcabello}.   This is different from the  EPR protocol or the BB84 protocol where %%@
only half of the EPR pairs or particles are used as keys.    We now study the efficiency %%@
of the scheme. The information-theoretic efficiency defined in Ref.\cite{rcabello} is
\begin{eqnarray}
\eta={b_s \over q_t+b_t},
\end{eqnarray}
where $b_s$ is the number of secret secret bit received by Bob, $q_t$ is the number of %%@
qubit used, $b_t$ is the number of classical bits exchanged between Alice and Bob during %%@
the QKD process. Here the classical bits used for evesdropping checking have been %%@
neglected. As has been discussed by Cabello\cite{rcabello}, in the BB84 protocol, %%@
$b_s=0.5$, $q_t=1$ and $b_t=1$. $b_t=1$ bit is used to indicate whether Alice and Bob use %%@
the same measuring apparatus. In this way, the efficiency of BB84 is 25\%. Similarly the  %%@
EPR protocol is 50\%. The protocol present here becomes 100\%. 

Another feature of our scheme is its high capacity since the 4 possible states of the EPR %%@
pair carry two bits of information whereas in the EPR scheme(BB84) each adopted EPR %%@
pair(particle) carries only one bit of information, in other words, $N$ adopted EPR pairs %%@
can send 2$N$ bits of key in our scheme.

 Townsend has introduced a protocol to distribute secret keys to multi-users over optical %%@
fiber networks\cite{rtownsend}. Townsend's protocol is a one-to-any protocol, where Alice %%@
acts as a single controller to establish and update a  distinct secret key with each %%@
network user. An any-to any protocl has been proposed to allow any two users to establish %%@
a secret key over an optical network by Phoenix et al\cite{rsimon}. The present scheme can %%@
be generalized to distribute secret keys to multiple legitimate users. The present is %%@
different from the previous two protocols in that the secret keys are common to all %%@
legitimate users. The procedure is given in the following. After Alice has sent the keys %%@
to Bob. Bob can create an EPR pair sequence which carries the raw keys. Then he uses this %%@
EPR pair sequence to another legitimate user, Clare, using the same procedure as before.  %%@
The key common to Alice, Bob and Clare are those Bell-basis measurement results that are %%@
not chosen to check eavesdropping. In this way, the protocol can be generalized to a %%@
multi-party common key distribution protocol. 

The implementation of the protocol proposed here require commensurate effort. Since it %%@
employs Bell state measurements, its practical implementation is difficult. Nevertheless, %%@
it is worth pointing that the operations employed here are all realizable in principle, %%@
for instance, the Bell-basis measurement was used in dense coding\cite{r7}. Recently %%@
complete Bell measurement has been realized in experiment\cite{rkim}. The sending of EPR %%@
partner particles was used in quantum clock synchronization\cite{r8}.  
Storage of light has been realized recently\cite{light1,light2}, and this may well serve %%@
to register the coming the  particle sequences and to store them. However, for a realistic %%@
implementation of the QKD scheme here, the efficiency of the Bell-basis detection and the %%@
length of time of photon storage need be enhanced.

  In conclusion, we propose a new QKD scheme, it is secure, efficient and has high %%@
capacity. 

Discussions with Dr Koashi is gratefully acknowledged. The authors are grateful for %%@
financial support from China National Natural Science Foundation, 
The Major State Basic Research Development Program contract no. G200077407, 
Hangtian
 Science Foundation, The Fok Ying Tung Science Foundation  for financial support.
We also thank the referee for helpful comments in improving the security of the protocol.

\noindent Figure caption

\noindent Fig.1. Schematic illustration of the new QKD scheme

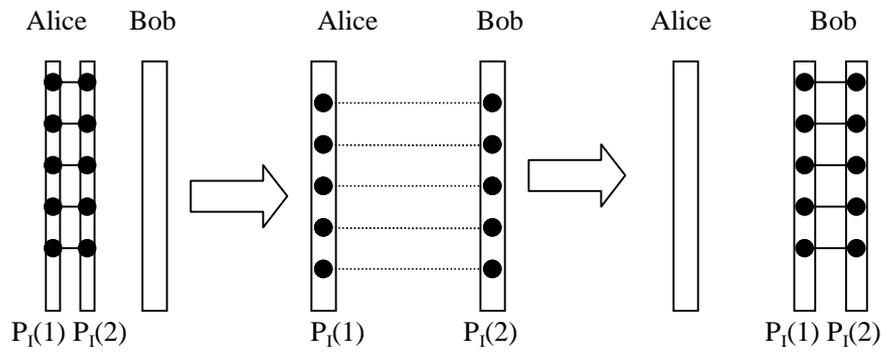
\end{document}